\documentclass{article}
\usepackage{graphicx} 
\usepackage[colorlinks=true,urlcolor=blue]{hyperref}
\usepackage{natbib}
\usepackage{listings}
\usepackage{authblk}

\title{\bf Spatial Data Science Languages: commonalities and needs}

\author[1]{Edzer Pebesma}
\author[2]{Martin Fleischmann}
\author[3]{Josiah Parry}
\author[4,13]{Jakub Nowosad}
\author[5]{Anita Graser}
\author[6]{Dewey Dunnington}
\author[7,12]{Maarten Pronk}
\author[8]{Rafael Schouten}
\author[9]{Robin Lovelace}
\author[10]{Marius Appel}
\author[11]{Lorena Abad}

\affil[1]{Institute for Geoinformatics, University of M\"{u}nster, Germany}
\affil[2]{Department of Social Geography and Regional Development, Charles University, Czechia}
\affil[3]{Environmental Systems Research Institute, Inc. (Esri)}
\affil[4]{Institute of Landscape Ecology, University of M\"{u}nster, Germany}
\affil[5]{Center for Digital Safety and Security, AIT Austrian Institute of Technology, Vienna, Austria}
\affil[6]{Wherobots, Inc.}
\affil[7]{Hydrology Software, Deltares, the Netherlands}
\affil[8]{Norwegian Institute for Nature Research (NINA), Oslo, Norway}
\affil[9]{Institute for Transport Studies, University of Leeds, UK}
\affil[10]{Department of Geodesy, Bochum University of Applied Sciences, Germany}
\affil[11]{Department of Geoinformatics - Z\_GIS, University of Salzburg, Austria}
\affil[12]{Urban Data Science, Delft University of Technology, the Netherlands}
\affil[13]{Institute of Geoecology and Geoinformation, Adam Mickiewicz University, Poznan, Poland}

\date{\today}

\begin{document}

\maketitle

\begin{abstract}
Recent workshops brought together several developers, educators and users of
software packages extending popular languages for spatial data handling,
with a primary focus on R, Python and Julia. Common challenges discussed
included handling of spatial or spatio-temporal support, geodetic
coordinates, in-memory vector data formats, data cubes, inter-package
dependencies, packaging upstream libraries, differences in habits or
conventions between the GIS and physical modelling communities, and
statistical models. The following set of insights have been formulated:
(i) considering software problems across data science language silos helps to understand and standardise analysis approaches, also outside the domain of formal standardisation bodies;
(ii) whether attribute variables have block or point support, and whether they are spatially intensive or extensive has consequences for permitted operations, and hence for software implementing those;
(iii) handling geometries on the sphere rather than on the flat plane requires modifications to the logic of {\em simple features},
(iv) managing communities and fostering diversity is a necessary, on-going effort, and
(v) tools for cross-language development need more attention and support.

\end{abstract}
\section{Introduction}
High-level open-source programming languages for data science enable users to combine
data management, interactive data analysis, and application development in
a single environment.
R \citep{R}, Python \citep{Python} and Julia \citep{Julia} dominate in this space. Spatial data science in these languages is the focus of this paper.
A growing focus
on `open science' principles and
reproducibility in scientific research have further increased the
demand for these languages \citep{hughes-noehrer_uk_2024}. A notable domain is
that of spatial data science, which focusses on datasets with explicit spatial locations. It is used in many applied fields, including climate, oceanography,
agriculture, ecology, biodiversity, geography, and mobility. Analysis tasks range from data collection and cleaning,
combining different datasets, exploring data, visualising it, predicting new observations to
statistical inference and identifying outliers or causalities.

R has the oldest roots of the three languages considered in this paper, dating back to the mid 1970s \citep{becker1986}.
R grew out of an experimental project to make an alternative
implementation of the statistical programming language S. In the 1997, R was open-sourced, and by
the early 2000s had largely replaced S and S-Plus in academia. Spatial R packages
have a history dating back to their development as S-Plus extensions. 
Early attempts to handle
spatial data were initiated by Roger Bivand, and reported
in \citet{classesmethods}. A coherent group of packages for spatial
statistics was described in \citet{asdar}, the first edition of
which appeared in 2008. Interface packages {\tt rgdal} \citep{rgdal}
and {\tt rgeos} were retired in 2023, as they had been
superseded by {\tt sf} \citep{rjsf,sds}, a package that
is standards-based and uses modern interfaces like {\tt Rcpp}
\citep{eddelbuettel2013seamless} and {\tt tidyverse} \citep{welcome}. 

Python was released in 1991 but its tooling for spatial data science
evolved much later.
The foundation of the ecosystem we use today can be linked to the work
of \citet{butler2005open} eventually resulting in the release of {\tt Shapely} \citep{Gillies_Shapely_2023}
as the Python bindings of {\tt GEOS} in 2007, {\tt Rtree} \citep{Butler_rtree_2023} as bindings of 
{\tt libspatialindex} in the same year and {\tt Fiona} \citep{Gillies_Fiona_2023} as bindings of {\tt GDAL} managing vector
geometries in 2008. Raster bindings through the {\tt rasterio} \citep{gillies_2019} package came 
five years later in 2013. The {\tt PROJ} bindings, known as {\tt pyproj} \citep{Snow_pyproj_2023} are around since 2006.
In parallel, the work of \citet{geoda} within the {\tt GeoDA/PySpace}
and that of \citet{rey_stars} in {\tt STARS} have merged and formed the Python Spatial 
Analytics Library ({\tt PySAL}) \citep{pysal}, at the time not tied to any of the 
bindings mentioned above. The modern ecosystem is much more connected
and oftentimes revolves around the {\tt GeoPandas} package \citep{geopandas}, released in 2013
to provide a common data frame-based interface to all {\tt Shapely}, {\tt Rtree}, and {\tt Fiona}.
In the recent years, {\tt PySAL} is progressing towards closer dependence
on the {\tt GeoPandas}-based ecosystem and deprecating its own geometry classes and
file interfaces. The world of raster data has been evolving largely in 
parallel based heavily on {\tt rasterio} and {\tt numpy} \citep{harris2020array}, with modern tools embracing
{\tt xarray} \citep{hoyer2017xarray} and the general Pangeo ecosystem.

Julia, introduced in 2012 and version 1.0 released in 2018, is a relatively young programming language \citep{Julia}.
Its spatial tools have also emerged recently.
Initially, efforts focused on wrapping C libraries like {\tt GDAL.jl} and {\tt LibGEOS.jl}, facilitated by Julia's call interface, automatic wrapper packages such as {\tt Clang.jl}, and its built-in package manager that supports binary dependencies.
These foundational packages have been abstracted to simplify working with C libraries.
More recently, higher-level abstractions like {\tt Rasters.jl} and {\tt GeoDataFrames.jl} have been developed, allowing for one-liner operations.
With many Julia packages incorporating (some form of) geometry, a spatial interface package called {\tt GeoInterface.jl} was created to enable spatial interoperability and conversion between different packages using traits.
The latest developments leverage Julia’s ability to compile high-performance native code, leading to the implementation of spatial algorithms directly in Julia, such as {\tt GeometryOps.jl}, rather than relying on the {\tt GEOS} wrapper.

Each programming language is characterised by communities
in which the exact boundary between developers and users is blurred:
users need to write commands to get something done, and many users
get into development sooner or later because they need to, and find
out they can.  Other programming languages like JavaScript, Java
and \href{https://github.com/georust/}{Rust} have also thriving communities working on spatial data,
but have a stronger division between users and developers. These
latter languages, as well as the older languages Fortran, C and
C++ have important components for spatial data analysis that are
often interfaced using binary language bindings and thus also
become useful to R, Python and Julia users.  
All three languages use a common representation for vector geometries defined by the simple feature access standard \citep{sfa}. Additionally, each language has infrastructure to read diverse data types such as raster (imagery), spatio-temporal array data (data cubes), and more traditional vector data formats as well as ways to interface with spatial databases, such as PostGIS and Spatialite with their built-in spatial analysis functions.

In September 2023, a hybrid Spatial Data Science across Languages (SDSL)
workshop was organised by Edzer Pebesma at the Institute for Geoinformatics
of the University of Münster. In September 2024, a second edition was organised by
Martin Fleischmann at the Charles University. These workshops served as a collaborative platform for
developers, educators, and users of software packages extending R, Python,
and Julia. The events were focused on addressing the challenges related to
the handling and analysis of spatial data. They fostered discussions among
experts in the field, leading to the identification of common challenges
and formulation of a set of recommendations, described in this document,
aimed at limiting potential pitfalls in spatial data analysis, cultivating 
communication between groups of developers and users, and 
enhancing the capabilities of spatial data science software.

\section{Common challenges}

\subsection{File formats, data connectivity, and in-memory representation}


A commonality across the spatial data science languages considered
is that each has a diverse and modular ecosystem of composable
packages/functions that can be mixed and matched to provide highly
customizable in-memory transformations compared to a traditional
desktop GIS-based processing workflow. Whereas desktop GIS-based
processing workflows typically combine tools that read files as
input and write files as output, spatial data science language
workflows commonly combine tools that transform data frames containing
one or more columns with a geometry data type.

Because many spatial data science workflows depend heavily on data
frame transformations, libraries in these languages are closely
related to the respective data frame libraries: {\tt GeoPandas} is an
extension of the pandas data frame library in Python; {\tt sf} is built
on R's built-in data.frame and provides methods for {\tt tidyverse}-style
transformations, and {\tt GeoDataFrames.jl} is built on top of the {\tt DataFrames.jl} package in Julia.
All of these add the concept of a "primary geometry column" and propagate its identity through as many transformations as possible.
Similarly, columnar geometry representations in R, Julia, and Python treat CRS as a column-level concept and propagate/validate its value wherever possible.

In-memory geometry column representation and examples of sharing between
packages is accomplished: In R, this is achieved with \verb|sf| and \verb|sfc| object classes, and some example
tools that leverage the ABI-stable (application binary interface) memory model are {\tt exactextractr},
{\tt stars}, {\tt geojsonsf}. In Julia, this is {\tt GeoInterface.jl}, or a set of
traits/generic methods that other packages can use to write generic
algorithms to read or transform an arbitrary geometry source. In Python,
this is a \verb|numpy| array of {\tt Shapely} objects, which are wrappers around
{\tt GEOS} geometries.

Since the development of geospatial support in R, Julia, and Python,
the diversity of tools available to work with data frames has exploded:
{\tt Polars}, {\tt DuckDB}, {\tt cuDF}, and Apache Arrow all offer some level of accelerated
data frame transformation in Python with varying levels of support in
R and Julia. As data size increases, users increasingly reach for these
tools and file formats such as Parquet that scale to support larger-scale spatial workflows.

Efforts to extend these libraries and formats to integrate
with the larger spatial ecosystem are underway: the OGC GeoParquet 1.1.0
specification was recently released \citep{Holmes2025opengeospatial} to extend the Parquet
format to mark columns as geometry and propagate the primary geometry
column, while Apache Parquet specification itself recently included native support for geometry and geography columns \citep{Dem2025apache}, which will be used by GeoParquet 2.0.0. The GeoArrow 0.1.0 specification \citep{geoarrow}, also recently released, includes both Apache Arrow extension types (enabling first-class
support for a geometry type in Arrow-based tools) and efficient
memory layouts compatible with DuckDB \citep{duckdb}, Polars \citep{polars}, cuDF \citep{rapids}, and Parquet.
DuckDB and cuDF both have geospatial extensions and work is underway
to improve connectivity and metadata propagation among geometry
representations in these libraries and geometry representations
in spatial data science ecosystems.

A commonality across the spatial data science languages considered is
that a typical workflow starts with reading data from an external
data source, like a file or database connection, compute, and write
results which may be a figure or again data written to some file.
A lot of interest has recently gone into efficient handling (reading,
analysing, writing) of large vector and raster data, in combination
with cloud computing. For vector data, "classic" file formats such
as Esri Shapefile or GeoPackage organize data record by record; for many
analytical purposes columnar storage formats such as GeoParquet can be considerably faster. In-memory data representations
such as GeoArrow, or those used by {\tt DuckDB} or {\tt Polars} can also speed up
processing and avoid copying of large amounts of data. Spatial
tiling such as done in FlatGeobuf or PMTiles may speed up reads of particular
regions. The growth and popularity of Apache Arrow as a common in-memory interchange 
representation is reflected in {\tt GDAL}, which optionally enabled reading of
vector file formats as an Arrow table in the version 3.6, followed by the
write support and inclusion of the GeoArrow specification two releases later.
This allows performant bulk I/O tailored to data frame representations of spatial
data.

Apache Arrow, together with its GeoArrow extension, has a potential to change the way
we move spatial data between different languages as Arrow can serve as an in-memory representation that can be accessed without copying from R, Python, Julia, Rust, C++, JavaScript, and others. Given  other languages like Rust or JavaScript
can also directly consume Arrow data, efficient support of Arrow should be a priority across all high-level languages.

For raster data and data cubes, formats such as
Zarr and cloud-optimized GeoTIFF (COG) are frequently used; these
allow for tiling, chunked reads, region reads, and low resolution
/ strided reads.  Such formats often require additional libraries
like {\tt BLOSC} and {\tt LERC} for compressing scientific (floating point) data.

\subsection{Inter-package dependencies and shared upstream libraries}

One of the powerful features of modern languages is that they can be
extended by packages that are developed and distributed by the community. Such package may reuse other packages,
allowing developers to build-on the work of others and focus on
things they know well. Important tasks like reading and writing
to external data formats, geometrical operations and conversions
or transformation of coordinates is typically left to external
libraries that are shared by a much larger open source community.

\subsubsection{Inter-package dependencies across languages} 

\begin{figure}
\begin{center}
\includegraphics[width=1.0\columnwidth]{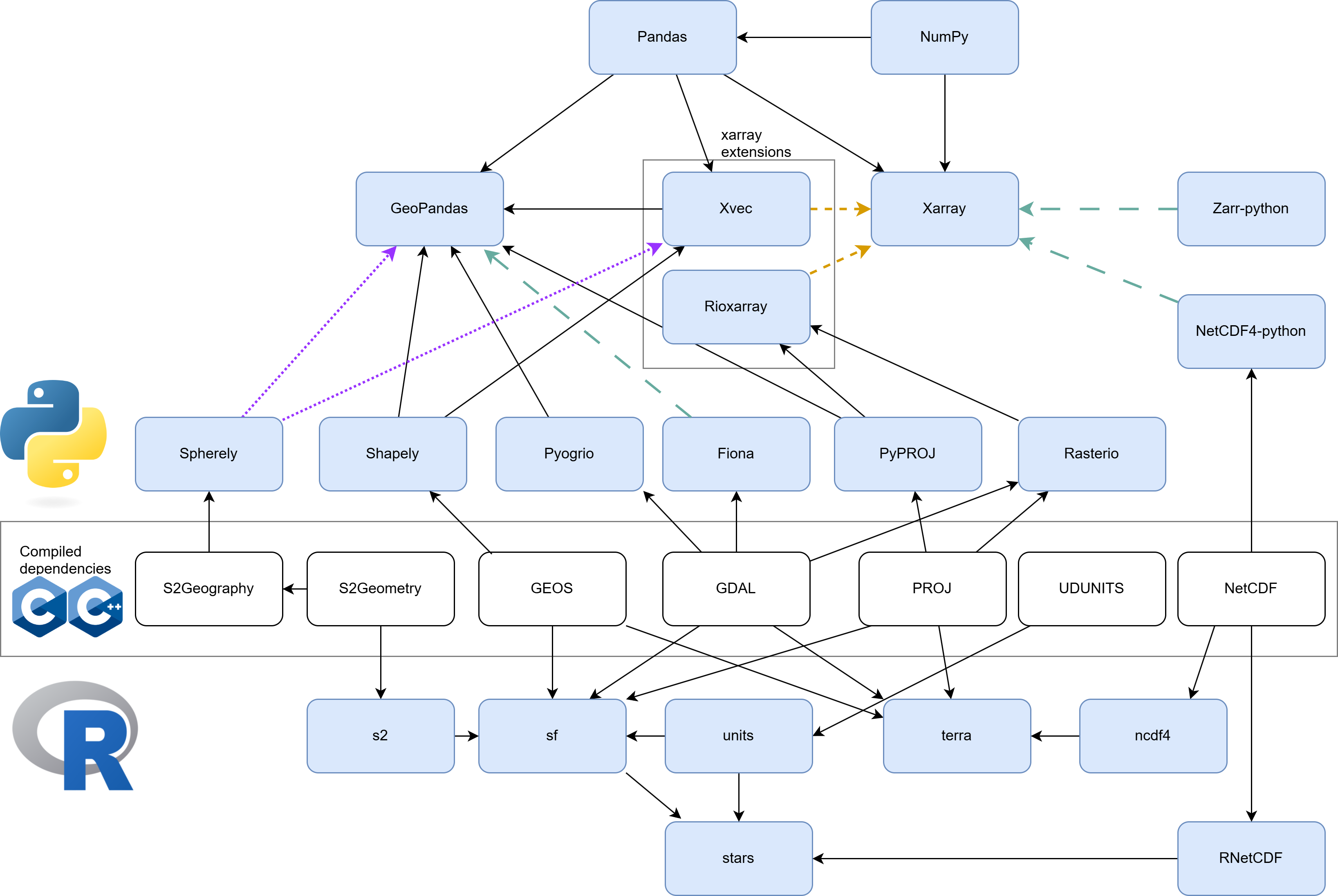}
\end{center}
\caption{Dependency of R and Python spatial packages on other libraries and external system requirements. Green long-dashed arrows indicate optional dependencies, purple dotted arrows indicate planned optional dependencies, while the orange dashed lines indicate optional dependency through the Xarray's extension mechanism.}
\label{fig:r_py_deps}
\end{figure}

A common feature of each language's support for spatial data is using existing low-level libraries for fundamental operations such as data import, export, and coordinate system transformation.
This approach is efficient because it avoids reinventing the wheel, but does mean that 'system dependencies' need to be managed somehow.
Each language needs to have access to {\tt GDAL} binaries, for example, and this is currently handled differently in each language. Figure \ref{fig:r_py_deps} gives an overview how the main geospatial R and Python packages depend on a common set up upstream C/C++ libraries.

In R, pre-built binary packages provided by the Comprehensive R Archive Network (CRAN) for Windows and Mac contain the system requirements: libraries such as {\tt GDAL}, {\tt PROJ}, {\tt UDUNITS} and {\tt S2Geometry} are compiled and statically linked to the shared library contained in the package. This leads to some duplication and inflexibility to upgrade system requirements by users, but has the advantage that binary packages, after installation, "just work". Linux users who install source packages must first install the system requirements, unless the package contains ("vendors") the complete library (as {\tt s2} and {\tt GEOS} do). Linux binary distributions of packages are also available, typically use dynamic linking and thus make implicit, hard requirements to installed versions; examples are r2u\footnote{\url{https://github.com/eddelbuettel/r2u}}, Posit Public Package Manager\footnote{\url{https://packagemanager.posit.co/)}} and bspm\footnote{\url{https://github.com/cran4linux/bspm}}.

In Python, {\tt GeoPandas} depends on {\tt Shapely} to provide {\tt GEOS} bindings, either on  {\tt Fiona} or  {\tt Pyogrio} to provide  {\tt GDAL} bindings and on  {\tt pyproj} to interface  {\tt PROJ}. In future, this list will also include  {\tt Spherely}, providing bindings to  {\tt S2Geometry}. The dependency on compiled C++ libraries and a resulting need for compiled distributions of Python packages directly depending on them may occasionally bring additional friction into the installation process. Historically, not all packages had released versions for all operating systems (often excluding Windows), which lead to the growth of independent community-led packaging efforts like {\tt conda-forge} \citep{conda_forge_community_2015_4774216} aiming to solve the packaging process by cross-compiling all packages in the ecosystem, ensuring binary compatibility and cross-platform support.

\begin{figure}
\begin{center}
\includegraphics[width=0.8\columnwidth]{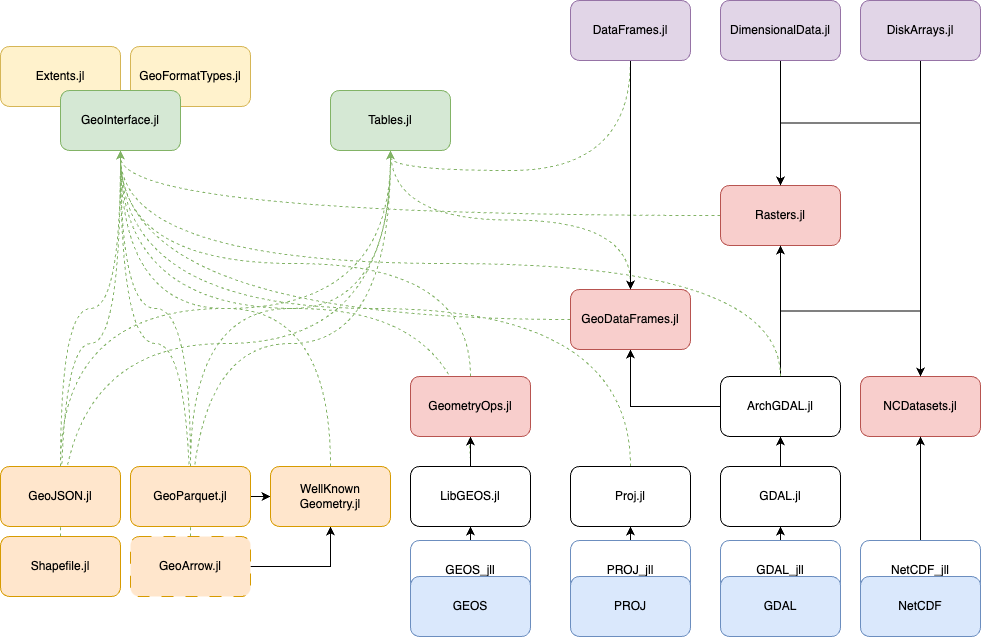}
\end{center}
\caption{Geospatial stack in Julia. Like other languages, the well known C++ libraries are wrapped (blue), resulting in the core packages of {\tt GeometryOps.jl}, {\tt GeoDataFrames.jl}, and {\tt Rasters.jl} (red). A number of packages implement native file formats (orange). Note that while the stack depends on other non-spatial Julia packages (purple), much of the interaction and indeed dependencies are replaced by the implementation of interfaces using traits (green). Only where native Julia types are not adequate, we created spatial versions of them (yellow).}
\label{fig:juliastack}
\end{figure}

Like the other languages, Julia wraps the C bindings of {\tt GDAL}, {\tt GEOS} and {\tt PROJ} libraries. However this is done in a generic way to produce versioned, shared binaries for the whole ecosystem. It does so by first cross-compiling all these projects, including all their (inter)dependencies as an artefact package (a so called .jll package, compared to normal .jl usage for a pure Julia package) for fourteen common platforms and application binary interfaces (ABIs). Packages like {\tt GDAL.jl} then depend on GDAL\_jll and wrap the C statements in Julia calls. Furthermore \emph{julian}---abstractions have been build on top, such as {\tt ArchGDAL.jl}, then {\tt Rasters.jl} and {\tt GeoDataFrames.jl} on top of that. In Julia it is easy to create new packages and extend functionality of other packages and types. This focus on composability has lead to an emphasis on interfaces, as seen in Figure~\ref{fig:juliastack}.

\subsection{Polygonal coverage}

In most of the cases discussed above, polygon data describe a
partitioning (tessellation) of the domain of interest, meaning that
no two polygons overlap and all area of interest is covered. This
refers to {\em polygon coverage} (for point support variables)
or {\em choropleth} (for block support variables). Knowing that a
dataset represents a valid polygonal coverage brings significant 
benefits to the subsequent processing as the algorithms depending
on overlay relations can avoid the need of testing the relationship
beyond the "touches" predicate. Furthermore, a valid coverage ensures
that vertices reflecting shared edges are present in both neighbouring
polygon, enabling fast coordinate-based construction of contiguity matrices
representing which polygon neighbours which. Yet, the simple
feature standard \citep{sfa} ignores this, and users are typically
faced with a set of polygon rings without any guarantee. Having
tools to verify overlaps, or to identify and remove small, unwanted
gaps or overlaps resulting from careless, topology-ignoring line
simplification would be convenient since geometrical operations become
more efficient when it is known that polygons do not overlap. This is currently
available only in a rather prototypical form. 

\subsubsection{Polygonal coverage across languages}

{\tt GEOS}, as of version 3.13.0, offers 
coverage validation and coverage gap finding to support its coverage-based algorithms
for simplification and union but not tooling for coverage enforcement and correction.
In Python, the package {\tt geoplanar} \citep{geoplanar}, 
covering most of the common planarity violations and their correction, attempts to fill this gap. We are not aware
of other attempts in R or Julia but users occasionally rely on JavaScript library 
{\tt mapshaper} \citep{mapshaper}, offering experimental tooling for topology cleaning,
or GRASS GIS \citep{GRASS_GIS_software}.

\subsection{Geodetic coordinates, spherical geometry}

Geodetic coordinates (also known as Geographic or
Geographical coordinates), degrees longitude and latitude describing
points on an ellipsoid or sphere, are fundamentally different
from Cartesian coordinates describing points on a flat surface.
For simplicity we will ignore here that the Earth's shape is better 
approximated by an
ellipsoid rather than a sphere, not because this is not important
but because the difference between sphere and ellipsoid is much
smaller than between a sphere and a flat surface. With the increasing
availability of global datasets, we also see an increase in datasets
with geodetic coordinates.

For geodetic coordinates with longitude $\lambda$ and latitude
$\phi$, the two points {\tt POINT($\lambda=0$ $\phi=90$)} and {\tt
POINT($\lambda=90$ $\phi=90$)} coincide, as do the two points {\tt
POINT($\lambda=-180$ $\phi=30$)} and {\tt POINT($\lambda=180$
$\phi=30$)}. For Cartesian coordinates, swapping the $x$ and
$y$ coordinate results in identical distance computations but 
that is not the case for geodetic coordinates. 
Geodetic coordinates measure angles and
have units in degrees, Cartesian coordinates measure distance and have
length units.

There is a long tradition in geospatial data analysis and GIS to treat
geodetic coordinates as if they are Cartesian \citep{chrisman}, which corresponds
to using the so-called plate carr\'{e}e projection, in which
$x = \lambda, y = \phi$.  As an example, the GeoJSON specification \citep{geojson}
prescribes to handle all geodetic coordinates as Cartesian in
the rectangle bound by {\tt POINT($x=-180$ $y=-90$)}
and {\tt POINT($x=180$ $y=90$)}. This may be helpful for
some problems, but not for others, and one wonders how many users
of GeoJSON are aware of these explicit intentions.

As an alternative, one might consider geodetic coordinates as
coordinates on a sphere. This, however, changes a number of fundamental 
things. For example, on a flat surface, any ring has an unambiguous inside and an
outside: only one of them has a finite surface. A sphere, by contrast, is a finite surface and any ring (polygon), therefore, divides the sphere's surface into two
finite parts. 

Similarly, on a flat surface, one can use the terms "clockwise" (CW)
and "counter-clockwise" (CCW) to indicate the winding order of the nodes
of a ring; on the sphere this is again ambiguous: if one takes the
ring formed by the equator, CW and CCW flip when one changes the side from which one
looks at it. 

Another issue is that straight lines are unambiguously straight on a flat
surface but, on a sphere, they are curved; the obvious choice is to use
{\em great circles arcs} (the shortest path over the sphere connecting
two points). Great circle arcs do not coincide to straight lines in
the plate carr\'{e}e view \citep{geojson}, except when they fall on a meridian or the equator. 

When using the simple
features standard to handle polygons on the sphere \citep{sfa},
one needs to additionally take care of:
\begin{itemize}
\item winding order: simple features require outer rings to have CCW,
inner rings (holes) to have CW winding, but this is not enforced
and most software does not require this (e.g. {\tt GEOS} produces geometry
with reversed windings). The options are either to define the inside
of a ring to be the area to the left when following nodes, or more pragmatically define
the inside as the smaller of the two polygons,
\item the \verb|POLYGON FULL|: the special polygon formed by the entire Earth's surface
as the negation of the empty set, \verb|POLYGON EMPTY|,
\item validity: polygons valid on plate carr\'{e}e are not generally
valid on the sphere, and vice versa \citep{sds}.
\end{itemize}

\subsubsection{Geodetic coordinates across languages} 


The R package {\tt sf} has, since its 1.0-0 release in June 2020
adopted the {\tt s2} package, which wraps the {\tt S2Geometry} C++
library to fully use spherical geometry operations {\em by default}
when provided coordinates are geodetic.

The Python package {\tt GeoPandas} aims to follow the behaviour of the R package
{\tt sf} and leverage the {\tt S2Geometry} C++ library when dealing with spherical
data. The work to create vectorized Python bindings to {\tt S2Geometry} enabling
this is currently ongoing within the {\tt Spherely} package and it is expected that
the first experimental support of spherical geometry in {\tt GeoPandas} will be released
in upcoming versions. So far, however, {\tt GeoPandas} shows a warning like in 
the example below where a GeoDataFrame with geographic coordinates is queried
for polygon areas. The "units"
of the values obtained are "squared degrees". For area, this is not
helpful as an area of 1\textdegree $\times$ 1\textdegree ~is over
12000 km$^2$ near the equator, whereas near the poles it is a bit
over 100 km$^2$. See the treatment of geodetic coordinates in {\tt GeoPandas} 0.14:

\begin{lstlisting}[breaklines]
>>> import geopandas
>>> countries = geopandas.read_file("countries/")
>>> countries.area
<stdin>:1: UserWarning: Geometry is in a geographic CRS. Results from 'area' are likely incorrect. Use 'GeoSeries.to_crs()' to re-project geometries to a projected CRS before this operation.

0         1.639511
1        76.301964
2         8.603984
3      1712.995228
4      1122.281921
          ...     
172       8.604719
173       1.479321
174       1.231641
175       0.639000
176      51.196106
Length: 177, dtype: float64
\end{lstlisting}

The warning message emitted by {\tt GeoPandas} is correct, but one
wonders how often data scientists seeing it will follow
the advice, as the warning does not suggest a concrete coordinate reference
system to use in \verb|to_crs()|. 
For area computations, one could choose an equal
area projections, transform the data, and recompute area. The
problem is harder for distance computations, or e.g. computing
buffers involving global or near-global datasets.  In such cases,
projection to Cartesian coordinates is not a great solution,
and computations on a spherical (or ellipsoidal) geometry are to
be preferred.

In Julia, most packages implicitly assume operations are based on Cartesian coordinates. Because not all geometries include a coordinate reference system (CRS) or can determine if the attached CRS is geodetic, the software cannot alert users, leaving them to verify the correctness of operations themselves. Some packages offer partial support for geodetic operations, such as calculating great circle distances, or provide workarounds like a \verb|cellsize| method to accurately compute raster areas.

Since the first SDSL workshop, the {\tt GeoInterface.jl} package has introduced a \verb|crstrait| trait, allowing packages to identify the CRS type and automatically select the appropriate method when available. This trait is currently being integrated across the ecosystem. Additionally, efforts are underway to wrap the {\tt S2Geometry} library in Julia, aiming to offer a comprehensive suite of geodetic operations.

Writing spatial data to dedicated file formats needs care if coordinate reference systems are not specified. Writing GeoPackage files with {\tt GDAL} ($\ge 3.9$) without specifying a CRS creates a file that, upon reading, specifies that coordinates are geodetic. A dedicated layer creation options (\verb|SRID="99999"|) or a dedicted WKT specification of an unknown {\em engineering CRS} specifies Cartesian coordinates, the default assumption in {\tt sf} and {\tt GeoPandas} for objects without specified CRS. GeoParquet also has a metadata flag that specifies whether line or polygon edges should be taken linear (Cartesian) or great circle arcs (geodetic). Some file formats (e.g., GeoPackage, GeoParquet, Parquet) interpret a missing CRS as having geodetic coordinates.

\subsection{Maps}
Among all statistical graphs, maps are compelling because it is so easy to relate to them. Yet, any map is a projection and creating code that maps unprojected data involves choosing defaults: should the projection preserve area, should Europe be at the center of the map, should North point up, should a graticule (a grid with lines following constant longitude or latitude) be added? 

Currently, the default plot method of R package {\tt sf} uses an equidistant rectangular projection, meaning that longitude and latitude are mapped linearly to $x$ and $y$, with unit scale (one unit distance in $x$ direction equals one unit distance in $y$ direction) at the center of the map. {\tt GeoPandas} applies the same treatment when dealing with unprojected data. Other software may choose for a plate carr\'{e}e projection (e.g., QGIS). More dedicated plotting packages such as the R {\tt tmap} choose a more sensible projection.  In particular for (near) global maps plate carr\'{e}e seems a poor choice, and some coordination between defaults maintained in different language environments is desirable. 

\subsection{Point versus block support}

The {\em support} of a data variable refers to the physical area
and/or temporal period with which an observed or computed data
value is associated (Table \ref{tab:examples}).  Spatial data
variables have an associated
geometry: a point, a line or a polygon (or a raster cell, which can
be seen as a polygon). In case the geometry is a point, the variable
is said to have {\em point support}. In reality, observation is
always associated with devices that have a non-zero size, but for
practical purposes we often treat small sizes as points. In all other
cases, the geometry has non-zero size (length, or area) and refers
to an infinite number of points. Data values of an attribute then
can refer either to the value of every point in the geometry, in
which case the variable has point support, or summarise a property
of all these points with a single value, in which case the
variable has {\em block support} (or line, area, or time period
support). As over space, variables have temporal point support when
associated with time instances and either temporal point or block
support when associated with time intervals.

Examples of a point support variables associated with polygons are
soil type, land use, and risk: all points in the polygon have the
associated value for soil type, land use, or risk category. Examples
of block support variables associated with polygons are population
count, population density, or standardised incidence rates for a
disease: reported values are counted or averaged over the polygon
and are not valid for every point it contains. Polygon boundaries of
a point support map indicate boundaries where the variable changes;
for block support variables they are often administrative boundaries,
and often carry no principled relation with the aggregated variable.

\begin{figure}
\begin{center}
\includegraphics[width=0.9\columnwidth]{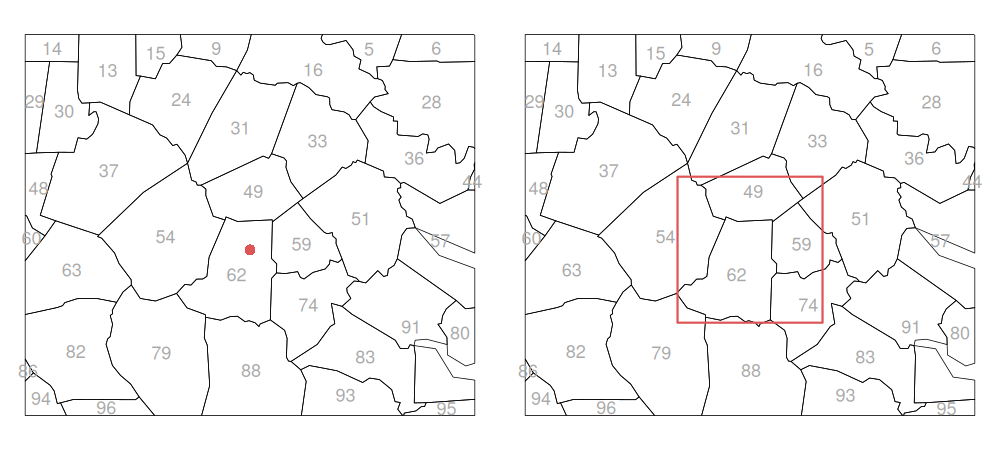}
\end{center}
\caption{Support of a variable: retrieving the values of the black
polygons for the red point or the red square needs knowledge whether
values (grey) are associated with {\em each} point in a
black polygon (point support) or whether they {\em summarise} properties of
{\em all} points in a polygon (block support). Vice versa, going from
red to black geometries involves the same problem.}
\label{fig:support}
\end{figure}

The support of a variable matters when data variables that are
associated with non-matching geometries are combined (figure
\ref{fig:support}). The disaggregation (figure \ref{fig:support}
left) or aggregation (figure \ref{fig:support} right) of point
support variables is trivial: as all values are known, there is
no uncertainty involved. Disaggregation and aggregation of a block
support variable is harder: the only thing we can assume is that at a finer
(e.g. point) support, the variable will not be constant throughout
the geometry. Auxiliary variables are needed to meaningfully model
this variation, and obtain downsampled or aggregated values. Examples
are downsampling of climate variables \citep{hewitson1996climate}, or dasymetric mapping of
population density \citep{mennis2009dasymetric}.

In addition, when disaggregating (downsampling) or aggregating
(upsampling) block
support variables, it needs to be known whether sums need to be
preserved, as with population counts, or whether averages need
to be preserved, as with population density or temperature.
When sums need preservation we call the variable {\em extensive},
when averages need preservation we call it {\em intensive}. These
properties may depend on whether considering space or time. As an
example, yearly CO$_2$ emissions of coal power plants have spatial
point support (power plant) and temporal block support (years),
and are spatially extensive but temporally intensive.

A typical spatial data science analysis starts with reading data
from an external source.  Ideally, this source should indicate
whether variables have point or block support in space and time,
and whether they are intensive or extensive. When this is the case,
the software used could choose reasonable defaults which methods to
use, or raise warnings if unexpected / potentially unsuitable
methods would be used. Also, the
software in turn could write back the data in a format documenting
the support and/or intensive/extensive flag.

Metadata containing support information is often scattered and
incomplete. GeoTIFFs can have a metadata flag, \verb|AREA_OR_POINT|,
which, when set, indicates whether raster cells refer to the value
of the cell area or at the cell center point. GeoPackage \citep{gpkg} or
FileGeoDatabase files can have a \verb|FieldDomain| value that indicates
a split policy and a merge policy. These two policies seem to
reflect whether a variable is extensive or intensive (table
\ref{tab:fielddomain}).

Redistricting (figure \ref{fig:support} right) always involves
both splitting and merging \citep{thomas, do, goodchild, sds}.
Table \ref{tab:examples} gives examples of extensive and intensive
variables for point or block support, associated with point, line
or area geometries. Note that while categorical variables can be 
conceptualised within this framework as (mostly) intensive, their
treatment in redistricting will be different. Typically, software
such as a Python package {\tt Tobler} \citep{eli_knaap_2024_14031618} reports 
proportions per each class in the output.

\begin{table}
\begin{center}
\begin{tabular}{|l|l|l|} \hline
Variable type & Split policy & Merge policy \\ \hline
extensive & geometry ratio & sum \\ 
intensive & duplicate & geometry weighted average \\ \hline
\end{tabular}
\end{center}
\caption{Values for the "split policy" and "merge policy" field domains for a variable that is spatially extensive or intensive }
\label{tab:fielddomain}
\end{table}

The NetCDF climate and forecast conventions \citep{cf} allow for 
specifying {\em bounds} on dimension variables (space and time) so
that particular areas (grid cells) or time periods are made explicit.
In addition, it uses the \verb|cell_methods| attribute for specifying
the aggregation {\em method}, indicating {\em how} a value was
aggregated over an area or period (table \ref{tab:cf}).

\begin{table}
\begin{tabular}{|l|l|l|p{7.5cm}|}\hline
Support & Dim &  Int/Ext & Example \\ \hline
Point & 0 & Int & temperature or air quality measured at a sensor \\
      &   & Ext & power plant CO$_2$ emission, capacity of a wind turbine \\
      & 1 & Int & Road surface reflectance \\
      &   & Ext & $\emptyset$ {\em (lines have an infinite number of points)} \\
      & 2 & Int & clay content of the soil \\
      &   & Ext & $\emptyset$ {\em (areas have an infinite number of points)} \\
Block & 0 & Int & $\emptyset$ {\em (block support implies non-point geometry)} \\ 
      &   & Ext & $\emptyset$ {\em (block support implies non-point geometry)} \\
      & 1 & Int & average width of a road, minimum width of a road \\
      &   & Ext & length of a road, duration to travel a road segment \\
      & 2 & Int & County average elevation, average temperature, county population density \\
      &   & Ext & County population total, county area \\ \hline
\end{tabular}
\caption{Examples of spatially extensive (Ext) and intensive (Int) variables having Block or 
Point support, for various geometry dimensions (Dim: 0 for points, 1 for lines, 2 for polygons); $\emptyset$ refers to situations where no examples are possible (reason in {\em italics})}
\label{tab:examples}
\end{table}

\begin{table}
\begin{tabular}{|l|p{4cm}|} \hline
CF encoding & Meaning \\ \hline
\verb|pressure:cell_methods = "time: point";| &  (temporal) point support \\
\verb|ppn:cell_methods = "time: sum";| & (temporal) block support, using sum of precipitation \\
\verb|maxtemp:cell_methods = "time: maximum";| & indicates (temporal) block support, giving the maximum over the period \\ \hline
\end{tabular}
		\caption{ CF-convention encoding of temporal support, and aggregation method}
		\label{tab:cf}
\end{table}

\subsubsection{Point versus block support across languages}

R package {\tt sf} deals to some extent with spatial support: {\tt sf} data.frames carry an attribute \verb|agr| (for attribute-geometry relationship) that for each non-geometry variable ("attribute") is either missing or has one of three values:
\begin{description}
\item[constant]: the value has point support, and has constant value for every point of the geometry
\item[aggregate]: the value has block support, and is the result of aggregating some quantity over the geometry
\item[identity]: the value identifies the geometry (e.g., a state name; this is point support)
\end{description}
When operations are carried out on an attribute where point support is assumed (e.g., area-weighted
interpolation, or extraction of polygon values a point geometries) and the \verb|agr| field is missing or \verb|aggregate|, a warning is raised. Spatial \verb|aggregate| or \verb|summary| methods can set the \verb|agr| flag of the attribute involved to \verb|aggregate|.

In Python, there is typically no built-in way to carry the information about point and block
support or of a type of a variable (intensive vs extensive). The responsibility in this case
is left to a user to determine the correct operation for each variable. The same situation is in Julia.

\subsection{Data cubes}

We use the word "data cubes" as synonymous with n-dimensional
arrays with data values (or tuples) arranged at the combinations of
a set of $n \ge 1$ dimension values. In our context, dimensions typically include
spatial and/or temporal variables. Dimensions may also be discrete,
e.g. a sequence of plant species when the array contains per species
abundance values or warehouses' locations where the array cells 
contain total sales (c.f. online analytical processing (OLAP) data cubes). 

A simple data cube example is a greyscale raster image, which
is a two-dimensional array. Multi-band (e.g. Red/Green/Blue) or
multi-time images are three-dimensional, while multi-band multi-time
images are four-dimensional. Non-raster data cubes can be time
series data for a set of (point) sensor locations, or time series
with averages for polygon geometries; these are two-dimensional.
The case of $n=1$ corresponds to one or more columns in a table.

A lot of large datasets lend themselves well to be represented as
data cubes, including Earth observation data, weather forecast or
reanalysis data, and climate model forecasts.

Common operations on data cubes include:

\begin{itemize}
\item creating regular data cubes from irregular or sparse ones (e.g., image collections collected at different times and having heterogeneous coordinate reference systems, or collections of trajectories),
\item interacting with the data cube at a lower spatial and/or temporal resolution, e.g. by
pyramids or overviews, where different ways of computing pyramids are possible,
\item analysing processes in space, time, or spacetime, e.g. by trend analysis or aggregation, identifying persistent spatial patterns or space-time interactions, and
\item creating machine learning models and classifying images.
\end{itemize}

\subsubsection{Data cubes across languages} 

R packages {\tt stars} and {\tt terra} provide two views on
representing data cubes, where {\tt stars} focuses on a generic
representation of (raster or vector) n-D cubes, and {\tt terra} on high performance
and scalability for stacks of raster layers (three-dimensional
data cubes). Both integrate well with vector data to extract
values at points or aggregate data cubes over polygons. 
Irregular image collections like time series of Sentinel-2 imagery (optionally fetched from a STAC catalog using {\tt rstac}, \cite{simoes}), can be transformed into regular four-dimensional data cubes using the {\tt gdalcubes} package \citep{appel}, supporting the use of image overviews at lower spatial resolution. 
A complete workflow from data cube creation to the application of machine learning methods for satellite image time series classification can be achieved using the {\tt sits} package \citep{camara}.
Support for moving features within data cubes is also being conceived in a package under development ({\tt post}). However, their complex nature still posses challenges for their implementation (see section \ref{trajectories}).

The Python package {\tt xarray} is used widely to analyse very large data cubes and is well integrated with {\tt dask} for parallel processing.
{\tt xarray} provides integration with spatial vector data (like R's {\tt stars} and {\tt terra}) through a series of
packages extending its functionality like the package {\tt xvec} for vector data cubes and zonal statistics.
To create data cubes from irregular image collections (with overlapping areas, different CRSes, etc.), {\tt StackSTAC} takes a STAC catalog and creates a regular data cube as a lazy four-dimensional Dask array. As a coordinating initiative for large-scale geoscientific applications in Python, Pangeo (pangeo.io) provides a community, a software ecosystem, and facilitates the deployment on computing infrastructure.

In Julia, {\tt Rasters.jl} and {\tt YAXArrays.jl} can handle arbitrary n-D cubes, 
building on the shared backends {\tt DimensionalData.jl} (for spatial indexing) 
and {\tt DiskArrays.jl} (for lazy chunk-based operation following Julias' 
`AbstractArray` interface). {\tt Rasters.jl} also handles stacks of mixed 
type and n-D layers, that can represent arbitrary NetCDF or grib files,
and integrates with all {\tt GeoInterface.jl} compatible vector data.
Cubes can be sampled using points in {\tt Rasters.jl}, and sampling using lines and polygons can be used with custom `Where` statements.
However, optimised spatial selectors have not yet been implemented.
{\tt DiskArrayEngine.jl} is in early  development, but extends {\tt DiskArrays.jl} 
for optimised out-of-core chunked operations leveraging {\tt Dagger.jl}, as
an analogue of {\tt xarray}/{\tt dask} in the Python ecosystem.

The openEO API \citep{openeo, mohr} is a higher level interface to
cloud-based processing systems similar to the closed source systems
Google Earth Engine \citep{earthengine} and Sentinel-Hub \citep{sentinelhub}.
Clients for openEO in R, Python, Julia, and JavaScript are available, and link to the respective package ecosystems. Examples of openEO-based processing systems are {\em openEO Platform} and the {Copernicus Data Space Ecosystem}.

\subsubsection{GIS and modelling community conventions}

\begin{table}
\begin{tabular}{|l|l|l|} \hline \hline
 & GIS  & Model (weather, climate, …) \\ \hline
Data types & Vector, raster &  Arrays, discrete global grids \\
Dynamic data & not of primary interest &  dominantly present \\
Height-varying data & ignored &  often present (atmosphere) \\
Coordinates & Cartesian  & Geodetic \\
Raster types & regular & regular, rectilinear or curvilinear \\
File formats & Esri Shapefile, GeoTIFF & NetCDF, Zarr, GRIB \\
Data standardisation & Open Geospatial Consortium & NetCDF CDL, CF Conventions \\
Data support & mostly ignored & CF: bounds, cell methods \\ \hline \hline
\end{tabular}
\caption{Comparison of the GIS and modelling communities' conventions}
\label{tab:conventions}
\end{table}

As opposed to the geospatial "GIS" community, the geospatial "modelling" communities comprises for instance weather, climate, oceanographic and geophysical modelling. To a large extent these communities create and share large datasets with predictions, forecasts, or reanalysis results, typically computed as a data cube, i.e. on a regular space and time discretization. For such datacubes the modelling community has largely converged on formats like NetCDF \citep{netcdf}, Zarr \citep{moore2023zarr}. The "climate and forecast" (CF) conventions \citep{eaton2003netcdf} constrain the very flexible NetCDF format to use specific names for dimensions, units, coordinate reference systems, and to express support and aggregation methods (bounds, cell methods). The data model underlying NetCDF and the CF conventions are also largely used in Zarr data cubes created by these communities. In a number of activities, the geospatial and modelling communities have attempted to find a common ground, these include the definition of geozarr, crs-in-netcdf, the {\tt GDAL} MultiDimimensional array C++ API, and the existence of NetCDF and Zarr drivers in {\tt GDAL}. A comparison of differences in conventions between the two communities is given in Table \ref{tab:conventions}.

\subsection{Trajectories} \label{trajectories}

While data cubes may cover time series of observations at points locations or for polygons, moving features and their trajectories represent another level of complexity in spatiotemporal data handling and analysis. Acknowledging this fact, the OGC has been working on a new standard -- beyond 
simple features -- called moving features \citep{mf} which encompasses standardized trajectory encodings in CSV, XML, and JSON, as well as a moving features access standard that specifies functions for movement data (similar to the spatial functions defined in the simple features access standard). 
In general, the OGC standard supports translation and/or rotation of moving features but not geometry deformation.

One of the main challenges is handling the temporal dimension and how discrete, linear, or step-wise interpolation of geometry and attribute information should be managed. A logical trajectory may consist of separate geometry and attribute time series whose time steps do not necessarily have to align. For example, a vehicle may be fitted with a GPS tracker recording at 1 Hertz and carry a sensor that records events at irregular time intervals. Trajectory data formats should support storage of this data, even if many analysis algorithms require that the data be synced before it can be analyzed further. 
Another challenge is the visualization of movement data. Common approaches include representing trajectories as directed lines (e.g. using arrow marker, gradients, or tapered lines) or using animation. However, most geospatial visualization libraries have no or very limited support for any of these options. 

\subsubsection{Trajectories across languages}  
The review paper of \cite{joo2020} lists 58 R packages created to deal with moving object or tracking data, with a large focus on movement ecology, and identified 11 packages with good or excellent documentation, mostly based on the CRAN Task View ``Handling and Analyzing Spatio-Temporal Data'' \citep{ctvspatiotemp}. A task view is an overview of R packages for a certain topic.
Some of the packages build on each other, however, a large number of packages are isolated even if having similar goals and methods.
From this review, the CRAN Task View ``Processing and Analysis of Tracking Data'' \citep{ctvtracking} was created to keep the focus on tracking data.
The tracking packages mainly depend on {\tt sf}, however, there are still a number of dependencies to {\tt sp}.


There are also numerous open source Python libraries -- most of which are based on either {\tt Pandas} or {\tt GeoPandas}, including: 
{\tt MovingPandas} \citep{graser2019}, 
{\tt traffic} \citep{Olive2019}, 
{\tt scikit-mobility} \citep{pappalardo2021},
{\tt TransBigData} \citep{Yu2022}
{\tt trackintel} \citep{MARTIN2023101938}.
For a full list, see \url{https://github.com/anitagraser/movement-analysis-tools/}.

As far as the authors are aware, there are no Julia packages for mobility analysis, although an effort to wrap the {\tt MEOS} library is currently underway.
Implementations discussed in the OGC Moving Features working group include the PostGIS extension {\tt MobilityDB} \citep{MobilityDBTODS2020} with its core functionality condensed in the C++ library {\tt MEOS} (Mobility Engine Open Source, a name inspired by {\tt GEOS}), and a Cesium-based viewer \citep{Kim2024aistairc}. The Moving Features JSON encoding (MF-JSON) is also supported by the Python library {\tt MovingPandas}.

\subsection{Statistical modelling}
One purpose common to using spatial data science languages is to fit some statistical model, diagnose it and carry out parameter inferences or create predictions. For doing so, typically data structures are created that specify the input or reflect the outcome of these steps. Inputs are for instance
\begin{itemize}
\item a model specification (e.g.~an R formula) specifying response, predictors, the type of their relationships, and random effects, or 
\item a point pattern with its observation window, or
\item neighbourhood lists or weights lists specifying which geographic features are related, and to which extent.
\end{itemize}
Output data structures may include fitted models, e.g. for
\begin{itemize}
\item predicting point density from covariates, or 
\item specifying residual spatial correlation in terms of a covariogram or semivariogram model,  or
\item spatial error or CAR models, specifying fitted regression coefficients and spatial correlation parameters.
\end{itemize}
As an example, \cite{bivandpackages} examined R and Python packages for areal data analysis, carrying out the same workflow in both environments and comparing the results.
Some data formats used in legacy software, e.g. .GAL files for neighbourhood lists, exist and can help transporting data structures from one environment to the other.

\subsection{Cross-language development}\label{sec:cross-lang-dev}

A common challenge that all SDSL attendees faced was using software from outside their main language.
This is difficult, not only because people tend to specialize on one language at a time, but also because package and environment development systems tend to be language-specific (CRAN does not host Python packages and PyPI does not host R packages, for example).
To overcome this challenge, there are two main options: cross-language environment managers and containerisation.
The {\tt pixi} environment manager, first \href{https://github.com/prefix-dev/pixi/releases}{released} in June 2023, aims to be a language-agnostic environment manager, with the ability to install R, Julia and spatial packages tested by SDSL attendees.

The current `gold standard' for creating and maintaining cross-language infrastructure --- for testing, development, and production --- is containerisation.
Docker containers can be developed that `ship' with all system dependencies.
They can be tagged to specific, immutable versions, providing long-term stability and a near-guarantee of reproducibility \citep{boettiger_introduction_2017, nuest}.
Efforts to provide Docker images explicitly for cross-language development are still rare.
Numerous projects maintain containers for cross-language spatial data science, including \href{https://darribas.org/gds_env/}{GDS} \citep{gds_env}, \href{https://rocker-project.org/}{Rocker} \citep{boettiger_introduction_2017, nust2020rockerverse}, \href{https://github.com/geocompx/docker}{geocompx}, and \href{https://github.com/b-data}{b-data}.

\section{Lessons learned and recommendations}

\subsection{Lessons learned}

The package ecosystems for R, Python and Julia are currently different in a
number of respects.  The R package {\tt sf} warns users if they are making 
implicit assumptions about line or polygon data having point support, and
has the ability to label data such that unnecessary warnings are not emitted.
It also uses geometrical operations by default on the sphere if
coordinates are not projected (geodetic), or warns if it has to
use operations in a Cartesian space for geodetic coordinates.
It raises errors if operations on objects with different
coordinate reference system are carried out, or bring them to a
common coordinate reference system as is done when plotted with {\tt ggplot2}.

Python's {\tt GeoPandas} package performs all operations on geodetic
coordinates in Cartesian space, while giving the user a warning that
values may be wrong and projection is recommended. Developments to
use a spherical geometry engine are in progress. There is no tooling
that would implicitly help with point and block support or differentiation
between intensive and extensive variables, although some {\tt PySAL} 
modules ask users to explicitly specify a variable type.

In Julia, most existing geometry interfaces either do not attach metadata such as a coordinate reference system, or have no method to distinguish between the different types of projection.
Avoiding combining objects with different coordinate reference systems and applying Cartesian operations to geodetic coordinates is the responsibility of the user, though automatic support for this is improving.

The three different languages have strongly different processes for
submitting and sharing packages. R packages distributed on CRAN
always need to work with other R packages released, and reverse
dependency checks are performed by CRAN on every new submission; changes
breaking other packages need to be communicated and clarified in
advance. 

Python has a complicated mechanism to check whether a release 
breaks any reverse dependency. Python packages have the ability to
require particular versions; package environments can be created
to create and maintain collections of packages with particular
versions. Furthermore, {\tt conda-forge} packaging system ensures compatibility of C++ components. 

Julia packages follow semantic versioning, and registered packages are required to set compatibility bounds for all declared dependencies. The combination of these factors means it is practically difficult to install packages that are incompatible. Doing reverse dependencies checks, and updating the compatibility bounds are the responsibility of the developer, but much of it is automated - for example, {\tt GDAL} tests {\tt ArchGDAL.jl}, {\tt GeoDataFrames.jl} and {\tt Rasters.jl} in continuous integration tests. Additionally, the Julia language infrastructure tests the entire ecosystem against all new release candidates to ensure breakages do not occur in relation to changes in the language itself.

\subsection{Recommendations}

\subsubsection{Open standards}

It is encouraging to see that involvement in the development of
open geospatial is no longer restricted to members of the Open
Geospatial Consortium (OGC), and takes place in issues of public
OGC GitHub repositories (for instance for GeoZarr and GeoParquet),
or even completely outside OGC communication channels (e.g. STAC,
GDAL, and openEO). OGC recently increased its individual and academic
membership fee from 550 USD to 2500 USD per year, making it more and more of
a business activity. Requiring more than one open source implementation
is still not a requirement for a standard to be adopted by OGC.

It would be convenient to have routines to verify polygons form a
coverage (i.e. are not overlapping), or to create non-overlapping
polygons from overlapping ones, in a way that all spatial
data science languages can profit from. A prototypical implementation
is found in R package {\tt sf}, with n-ary intersection and
difference, and in Python package {\tt geoplanar} \citep{geoplanar} providing a tooling to
fix common planarity issues. 

The increased diversity in data frame libraries, increased diversity
in geospatial tooling, and increasing data size
have highlighted a need for connectivity that extends beyond the
columnar memory model implemented separately in each of the spatial
data science languages. Wider adoption of GeoParquet as a more efficient
file format for whole-file read/write and wider adoption of GeoArrow as
a common metadata standard and memory model may help address these
challenges.

\subsubsection{Field domains}

Splitting spatially extensive block support variables to point
geometries should at all times be avoided: one would need an infinite
number of zero valued points that should sum to the polygon sum. With
discrete computers this cannot be realised, and software should warn
against attempts doing so. Intensive and extensive variables should
both refer to quantitative variables, otherwise sums, proportions
and averages do not make sense.  Split rules can only apply to
variables with non-point geometries.

As table \ref{tab:fielddomain} reveals, split and merge policies
follow from a variable being spatially extensive or intensive,
and hence one of them is obsolete. The question remains what to do
when a data source has two contradictory values set, e.g. split
policy "duplicate" and merge policy "sum". A less ambiguous
approach would be to have a single field domain called \verb|is_spatially_extensive|
with a boolean value.

\subsubsection{Geodetic coordinates, spherical geometry}

For a consistent handling of simple feature geometries on a spherical
geometry, one has to augment the simple feature standard in the following
way:
\begin{itemize}
\item in addition to the \verb|POLYGON EMPTY| one needs the \verb|POLYGON FULL|
to define the polygon formed by the entire surface of the sphere. In addition
to this WKT definition one would need a WKB representation of this; the value
used for this by the {\tt S2Geometry} library, WKB of \verb|POLYGON ((0 -90, 0 -90))|
would be a pragmatic choice. R package {\tt sf} now supports the \verb|POLYGON FULL|.
\item straight lines on the sphere are not straight: the obvious
choice is to use shortest arcs over the sphere ("great circle
distance"). When importing GeoJSON files, one may need to add nodes
to longer line segments on plate carr\'{e}e, to follow the GeoJSON
definition of a straight line.
\item one needs to define winding order of polygon rings, e.g. the
way {\tt S2Geometry} does this: the area left of the lines when connecting
nodes in order of appearance is the inside. When importing polygons,
a pragmatic default, also taken by BigQuery GIS \citep{bigquerygis}
is to define the inner side of a ring as the smaller of the two
areas, but full user control is needed to work e.g. with the polygon of the oceans.
\end{itemize}

\subsubsection{Community management}

R, being the oldest of the spatial data science languages, has now
finished a full development circle where crucial infrastructure
packages ({\tt rgdal} and {\tt rgeos}) have been developed, used,
maintained, and retired. Retirement involved a process taking more than
two years, informing all reverse dependencies when this would be
coming and how they could adapt to use modern alternatives.

The Python community is trying to coordinate the development efforts
potentially refactor core packages to support current needs to avoid
the need for retirement (which is also not entirely possible due to the design
of Python packaging systems) and migration.
The clear example is the recent
merge of the {\tt PyGEOS} project into {\tt Shapely}, resulting in {\tt Shapely} 2.0 release with
completely revised interface to {\tt GEOS}.

Julia is much earlier in the cycle, and while it has the benefit
of learning from the work of the R and Python ecosystems, has a relatively
small developer community, largely focused on high-performance or
big-data modelling applications.

The developer communities would benefit from a stronger diversity,
and one way to increase that is to create a welcoming culture for
newcomers, for comments, to adopt good code of conducts, and to
follow the experience collected in organisations like rOpenSci and pyOpenSci.
Several efforts towards this goal have already started as precursors and within the SDSL community. 
To foster support, guidance and to increase interaction between community members, discord channels are available both for the geocomp{\it x} project \footnote{\url{https://geocompx.org/}} and the SDSL community \footnote{\url{https://spatial-data-science.github.io/}}. 
The geocomp{\it x} discord is a open forum to seek support, ask questions, and showcase exciting work.
The SDSL discord serves as a tool to foster discussions, both among developers and users on spatial data science concepts, methods, packages and programming languages.

\subsubsection{Cross-language collaboration and infrastructure}

Communities and infrastructure for cross-language collaboration are in their infancy.
However, an increasing number of projects are using multiple languages for spatial data science for research, testing and deployment of services in production.
This creates a need for more cross-language events like the SDSL initiative that formed a basis for this paper.

One of the findings from the SDSL sessions is that cross-language infrastructure is in its infancy.
Projects such as `Rocker' and `Pixi', mentioned in Section~\ref{sec:cross-lang-dev}, have made a start in this direction, but they are still focussed on one language with support for other languages being add-ons, rather than core features.
We recommend that more work is done to provide both the technical infrastructure and social environments for constructive cross-language work for spatial data science developers, users and educators.

The geocomp{\it x} project is ``a community-driven effort to provide resources for learning and teaching about geocomputation in multiple programming languages''. Three books on the topic of geocomputation, one focusing on R\footnote{\url{https://r.geocompx.org/}} \citep{lovelace_geocomputation_2025}, one on Python\footnote{\url{https://py.geocompx.org/}} \citep{dorman_geocomputation_2025}, one on Julia\footnote{\url{https://jl.geocompx.org/}}, are available online through this project. The {\em Spatial Data Science} book \citep{sds} is also fully available online\footnote{\url{https://r-spatial.org/book/}}, a second edition is evolving with subtitle ``With applications in R and Python''\footnote{\url{https://r-spatial.org/python/}}, and contains tabbed code sections with R and Python tabs. The Quarto\footnote{\url{https://quarto.org/}} publishing system can be used to create books or websites from notebook-style markdown documents that contains multiple data science languages. Jupyter \citep{jupyter} notebooks can also handle multiple languages, but not in single document, as Quarto does.

\subsubsection{Future events}

The general impression of the attendees of the workshops on
{\em Spatial Data Science Languages} was very positive, as many
developers and users usually work in relative isolation, with asynchronous
on-line communication. It was agreed to repeat this event,
and to combine it then with a hackathon in order to do practical
work, including testing, comparison, and further interoperability
experiments.
Not all aspects of spatial data science have been considered yet and 
future events should also address other relevant domains such as spatial network analysis,
web-mapping and visualization, among others.

\section*{Acknowledgements}
The authors are grateful to all participants who contributed to discussions during the two spatial data science languages workshops, and in particular to Yomna Eid, who was instrumental in the organisation of the workshops.

\bibliography{main}
\end{document}